\documentclass[prc,preprint,showpacs,preprintnumbers,amsmath,amssymb,superscriptaddress]{revtex4-1}
\usepackage{graphicx}
\usepackage{dcolumn}
\usepackage{bm}
\usepackage{floatrow}
\usepackage[caption=false]{subfig}
\usepackage{booktabs}
\usepackage{multirow}
\usepackage{siunitx}
\usepackage{subfig}
\usepackage{longtable}
\usepackage{footnote}

\begin{document}
\preprint{}

\title{Structure of $^{38}$Cl and the quest for a comprehensive shell model interaction.}

 \author{R. S. Lubna}
 \affiliation{Department of Physics, Florida State University, Tallahassee, FL 32306, USA}
 \author{K. Kravvaris}
 \affiliation{Department of Physics, Florida State University, Tallahassee, FL 32306, USA}
 \affiliation{Lawrence Livermore National Laboratory, Livermore, CA 94550, USA}
 \author{S. L. Tabor}
 \affiliation{Department of Physics, Florida State University, Tallahassee, FL 32306, USA}
 \author{Vandana Tripathi}
 \affiliation{Department of Physics, Florida State University, Tallahassee, FL 32306, USA}
 \author{A. Volya}
 \affiliation{Department of Physics, Florida State University, Tallahassee, FL 32306, USA}
 \author{E. Rubino}
 \affiliation{Department of Physics, Florida State University, Tallahassee, FL 32306, USA}
 \author{J. M. Allmond}
 \affiliation{Physics Division, Oak Ridge National Laboratory, Oak Ridge, Tennessee 37831, USA}
 \author{B. Abromeit} 
 \affiliation{Department of Physics, Florida State University, Tallahassee, FL 32306, USA}
 \author{L. T. Baby}
 \affiliation{Department of Physics, Florida State University, Tallahassee, FL 32306, USA}
 \author{T. C. Hensley}
 \affiliation{Department of Physics, Florida State University, Tallahassee, FL 32306, USA}

\date{\today}

\begin{abstract}

The higher-spin structure of $^{38}$Cl ($N = 21$) was investigated following the $^{26}$Mg($^{14}$C, $pn$) reaction at 30 and 37 MeV beam energies.  The outgoing protons were detected in an $E- \Delta E$ Si telescope placed at 0$^\circ$ close to the target with a Ta beam stopper between the target and telescope.  Multiple $\gamma$ rays were detected in time coincidence with the protons using an enhanced version of the FSU $\gamma$ detection array.  The level scheme was extended up to 8420 keV with a likely spin of 10 $\hbar$.  A new multishell interaction was developed guided by the experimental information. This FSU interaction was built by fitting to the energies of 270 experimental states from $^{13}$C to $^{51}$Ti.  Calculations using the FSU interaction reproduce observed properties of $^{38}$Cl rather well, including the spectroscopic factors.  The interaction has been successfully used to interpret the $1p1h$ and $2p2h$ configurations in some nearby nuclei as well.

\end{abstract}

\maketitle

\section{Introduction}

The evolution of shell structures, specially with increasing proton-neutron imbalance, can provide valuable insights into the finite many-body problem. The exploration of an exotic region in the chart of nuclides with extreme $N/Z$ ratios, the so called ``island of inversion'' region, has cast doubt on the persistence of the classical magic numbers and revealed the fragility of the shell gaps that lead to the magic numbers. The nuclei with $Z = 10 \sim 12$ and  $N \approx$ 20 have been found to have ground states dominated by the intruder configurations from the upper $fp$ shell orbitals \cite{thibault, MOTOBAYASHI, huber, detraz, GUILLEMAUD, YANAGISAWA} and the anomalous property was interpreted as the reduction of the $N = 20$ shell gap. The immediate question emerged after the revelation of the shell gap reduction was how does this change happen along an isotopic or isotonic chain. Explaining this trend of shell evolution or structural evolution has been a great challenge to the nuclear structure models. The monopole parts of the shell model Hamiltonian have long been recognized to play the major role in the evolution of shell structure. While some models \cite{sdpf-u, sdpf-m} are very successful in explaining the very neutron rich $sd$ shell isotopes, they were unable to explain the intruder states of some nuclei within the same isotopic chain \cite{wbp-m}, or simply some other $sd$ nuclei which are not even very neutron rich \cite{heavyIon1}; meaning that their monopoles are not well determined to explain the shell gap evolution. This demonstrates the need of a more comprehensive shell model treatment for the intruder states of the $sd$-shell nuclei which are sensitive to the shell gaps and, hence, very informative to describe the shell gap evolution. \\


The current experimental investigation focuses on the structure of moderately neutron rich $^{38}$Cl with $Z=17$ and $N=21$ having the valence protons in the $sd$ shell and one valence neutron within the $fp$ shell. Both normal and intruder states of $^{38}$Cl are valuable to understand the $N=20$ shell gap evolution.  This nucleus has long been recognized as providing a window into the interactions between $\pi 0d_{3/2}$ and $\nu f_{7/2}$ nucleons, since the first 4 states ($2^-$, $5^-$, $3^-$, and $4^-$) comprise the expected multiplet \cite{transfeReac2}. More recently, a study of $\gamma$ decay of $^{38}$Cl following production in a binary grazing reaction revealed a cluster of 3 energy states around 3.5 MeV, one of which was suggested to involve an $f_{7/2}$ proton coupled to maximum spin ($7^+$) with the $f_{7/2}$ neutron \cite{heavyIon1}.  Other investigations studying $^{38}$Cl, including $\beta$ decay, transfer reactions, and neutron capture \cite{betaDecay1, betaDecay2, transfeReac1, transfeReac2, transfeReac3, nuRes1, nuRes2, heavyIon2}, have not proposed any state of spin greater than $5\, \hbar$. The availability of a $^{14}$C beam in conjunction with our $\gamma$-detector array has allowed further exploration of the higher-spin structure of $^{38}$Cl to elucidate the role of excitations across the $N = 20$ shell.  \\

Comparison with microscopic structure models has proved very fruitful in the past in interpreting level schemes, but were limited by the need to adjust the $N = 20$ shell gap for cross-shell excitations for different nuclei \cite{Lubna, wbp-b, wbp-m, wbp-a, heavyIon1}.  Therefore, we decided to develop a microscopic effective interaction based on fitting the shell model cross-shell interaction matrix elements over a wide range of particle-hole states in nuclei across the $sd$ shell and beyond. The main focus was to tune the monopole terms across the shell gaps, $N = 8$ and $N =20$ to reproduce the observed experimental data. The valence space of the new FSU interaction comprises the $spsdpf$ model space, compatible to the normal and intruder states of the $sd$ shell isotopes. The shell model calculations using FSU interaction have been performed in this work to better understand the structure of $^{38}$Cl and some nearby isotopes.\\


\section{Experimental Details and Procedures}

The experiment was performed at the John D. Fox laboratory at Florida State University using the $^{26}$Mg($^{14}$C, $pn$)$^{38}$Cl fusion-evaporation reaction. Two beam energies, 30 and 37 MeV, were delivered by the FSU Tandem accelerator. A 770-$\mu g/ cm^2$ self-supporting target, isotopically enriched to 99.6$\%$ in $^{26}$Mg and backed by a 25-$\mu$m thick tantalum stopper, was used. Coincident $\gamma$  rays were detected with the FSU $\gamma$-detector array consisting of three ``clover'' detectors, each having four HPGe crystals, and three single-crystal Ge detectors. For this experiment three more unsuppressed clover detectors from Oak Ridge National Laboratory were added to the array. Four clover detectors along with a single crystal one were paced at 90$^0$, two clovers at 135$^0$ and the rest two single crystal detectors were place at 45$^0$ angles.  The emitted charged particles were detected and identified using an $E-\Delta E$ silicon telescope with thicknesses of 1.0 mm and 0.1 mm,  respectively. The particle telescope was placed at $0^0$ with respect to the beam direction. Protons, deuterons, tritons, and alphas were clearly separated in the $E-\Delta E$ telescope, and the proton gate was chosen in order to select $\gamma$-rays in $^{38}$Cl. \\

The energy and efficiency calibrations of the $\gamma$ detection system were performed using a calibrated $^{152}$Eu source from NIST \cite{NIST} and a shorter-lived $^{56}$Co source made at FSU. Doppler corrections of the $\gamma$-ray energies were carried out with $\beta \,(v/c)$ values of 0.0172 and 0.0190 for the beam energies of 30 and 37 MeV, respectively. Two symmetric $E_\gamma$ - $E_\gamma$ matrices with and without the detection of one proton were created for building the level scheme. \\

To help with spin assignments for the newly observed states, a Directional Correlation of the $\gamma$ rays de-exciting Oriented states (DCO) analysis \cite{KRAMERDCO} was performed. For that, an asymmetric $E_\gamma$ - $E_\gamma$ square matrix was created by using the detectors at 90$^0$ on one axis and those at 135$^0$ on the other. The experimental DCO ratio used here is,

\begin{equation}
R_{DCO}=\frac{I_{\gamma 2}(at\, 135^0\, gated\, by\, \gamma_1\, at \,90^0)}{I_{\gamma 2}(at\, 90^0\, gated\, by\, \gamma_1\, at\, 135^0)}
\end{equation}\\

where, $I_{\gamma 2}$ denotes the intensity of the $\gamma$ transition whose multipolarity is to be measured and $\gamma _1$ is the gating transition. If the gating transition is a pure stretched dipole type, the $R_{DCO}\sim$1 for a pure dipole transition and $R_{DCO}\sim$2 for a pure quadrupole transition. On the other hand, if the gating transition is a pure stretched quadrupole transition, $R_{DCO}\sim$1 for a pure quadrupole transition and $R_{DCO}\sim$0.5 for  a pure dipole transition. The systematic errors for the DCO ratio method are relatively small since $\gamma _1$ and $\gamma _2$ are measured in coincidence. In order to test the correlation of the emitted $\gamma$ rays in the DCO ratio analysis, we have measured the $R_{DCO}$ of known transitions in $^{38}$Ar present in the same data set. The transitions at energies 670 and 1822 keV are reported as dipole \cite{nndc}, which agrees very well with our measurements, as shown in Fig. \ref{fig:RDCO}.\\

In conjunction with the angular correlation analysis, linear polarization measurements of the Compton-scattered $\gamma$ rays between crystals of the clover detectors were carried out for the parity assignment to many of the states. This measurement was conducted by using the four clover detectors at $90^0$ where the polarization is maximum. A clover detector consisting of four crystals performs as four Compton polarimeters where each crystal acts as a scatterer and the two adjacent crystals act as absorbers. The electric or magnetic nature of a transition can be extracted from the polarization asymmetry of the $\gamma$ ray when it hits one crystal of a clover detector and Compton scatters to an adjacent crystal in the same clover. The experimental polarization asymmetry of the Compton scattered $\gamma$ ray is defined as \cite{polFormula}

\begin{equation}
\label{eqn:pol}
\Delta = \frac{a(E_{\gamma})N_\perp - N_{\parallel}}{a(E_{\gamma})N_\perp +N_{\parallel}}
\end{equation}\\

In the above expression, $N_{\perp}$ and $N_{\parallel}$ are the numbers of Compton scattered photons of a given $\gamma$ energy along the perpendicular and parallel directions measured with respect to the beam direction. The scaling factor $a(E_\gamma)$ is defined as

\begin{equation}
a(E_\gamma)=\frac{N_\parallel (unpolarized)}{N_\perp (unpolarized)}
\end{equation}\\

The factor $a(E_\gamma)$ was measured in the current analysis by using $^{152}$Eu and $^{56}$Co sources over an energy range of 344 to 2597 keV and was found to be 1.01 (1). Also, no energy dependence was observed in $a(E_\gamma)$. Hence, for a pure electric transition, $\Delta > 0$ because an electric transition prefers scattering perpendicular to the beam axis in this type of experiment. On the other hand, $\Delta < 0$ for a pure magnetic transition because it favors parallel scattering along the beam axis \cite{KN_ZPhys} \cite{KN_nature}. The polarization sensitivity was measured using the $p+^{24}$Mg resonance reaction at $E_p = 6035$ keV, forming a $(9/2^+)$ state in $^{25}$Al at 8077 keV \cite{polRes}.  This state decays back by proton emission partially to the 4123-keV $4^+$ state in $^{24}$Mg.  We have measured the polarization asymmetry of the 2754 (4$^+ \rightarrow$ 2$^+$) and 1369 (2$^+ \rightarrow$ 0$^+$) keV $\gamma$ decay cascades, which are found to be consistent with their electric nature (Fig. \ref{fig:polarization}) as reported in the literature \cite{nndc}. A further sensitivity test was performed by examining the two transitions of $^{38}$Ar at 670-(M-type) and 1822-(E-type) keV energies which were found (Fig. \ref{fig:polarization}) to be in good agreement with the literature \cite{nndc}. All of the measured polarization asymmetries for the $\gamma$ rays of $^{38}$Cl along with the test cases are shown in Fig. \ref{fig:polarization}.\\ 


\section{Analyses and Results}

\begin{figure}[h!]
\centerline{\includegraphics[scale=0.7]{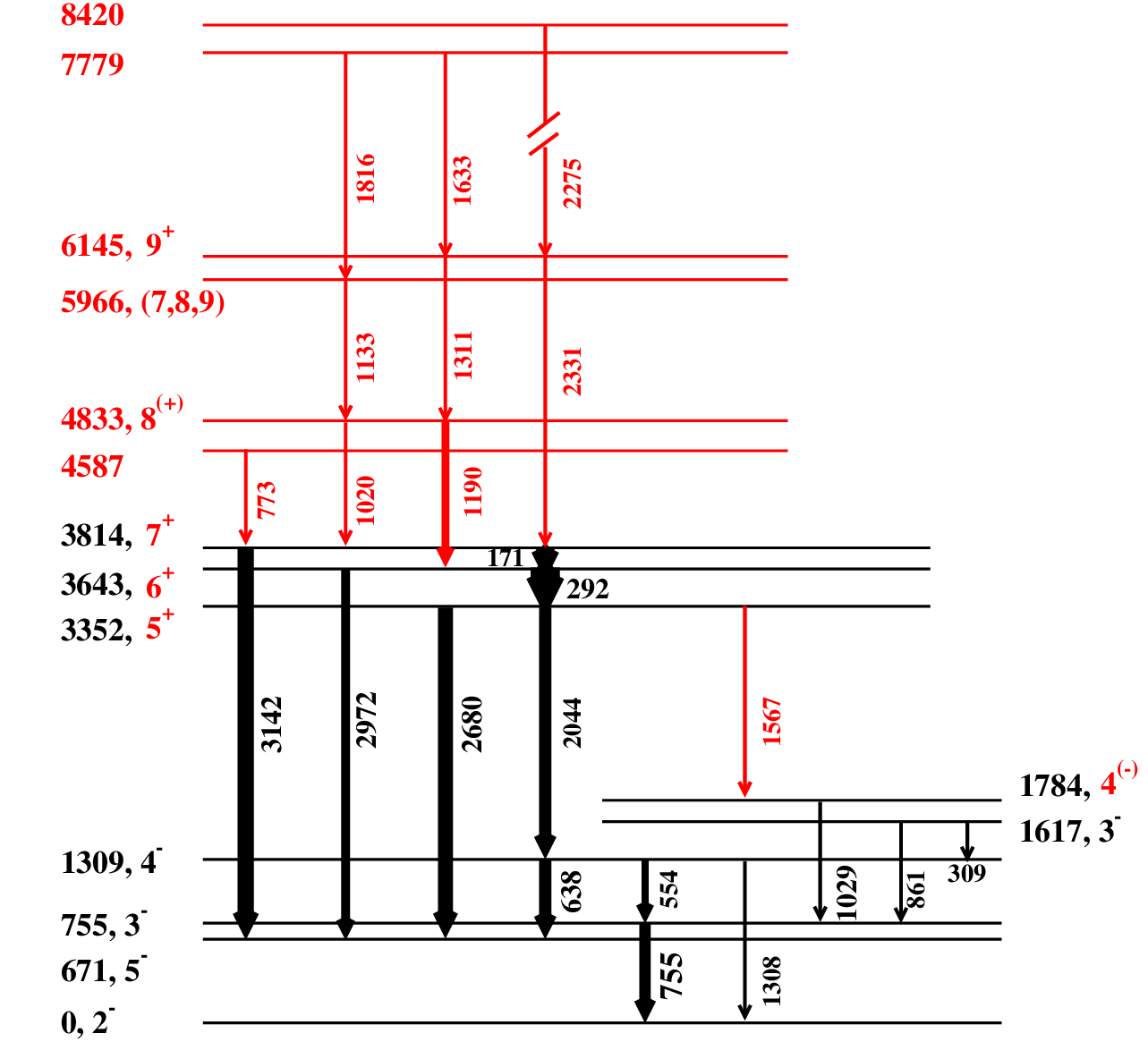}}
\caption{The level scheme of $^{38}$Cl established from the current analysis. States and transitions in red (gray) have been observed for the first time. The spins in the figure are from the current analysis and also from the literature \cite{nndc}. The arrow widths are proportional to the relative intensities of the $\gamma$-ray transitions normalized to the 292 keV transition. Those with less than $5\%$ intensity are drawn with the same width. }
\label{fig:38_level}
\end{figure}

\begin{figure}[h!]
\centerline{\includegraphics[scale=0.7]{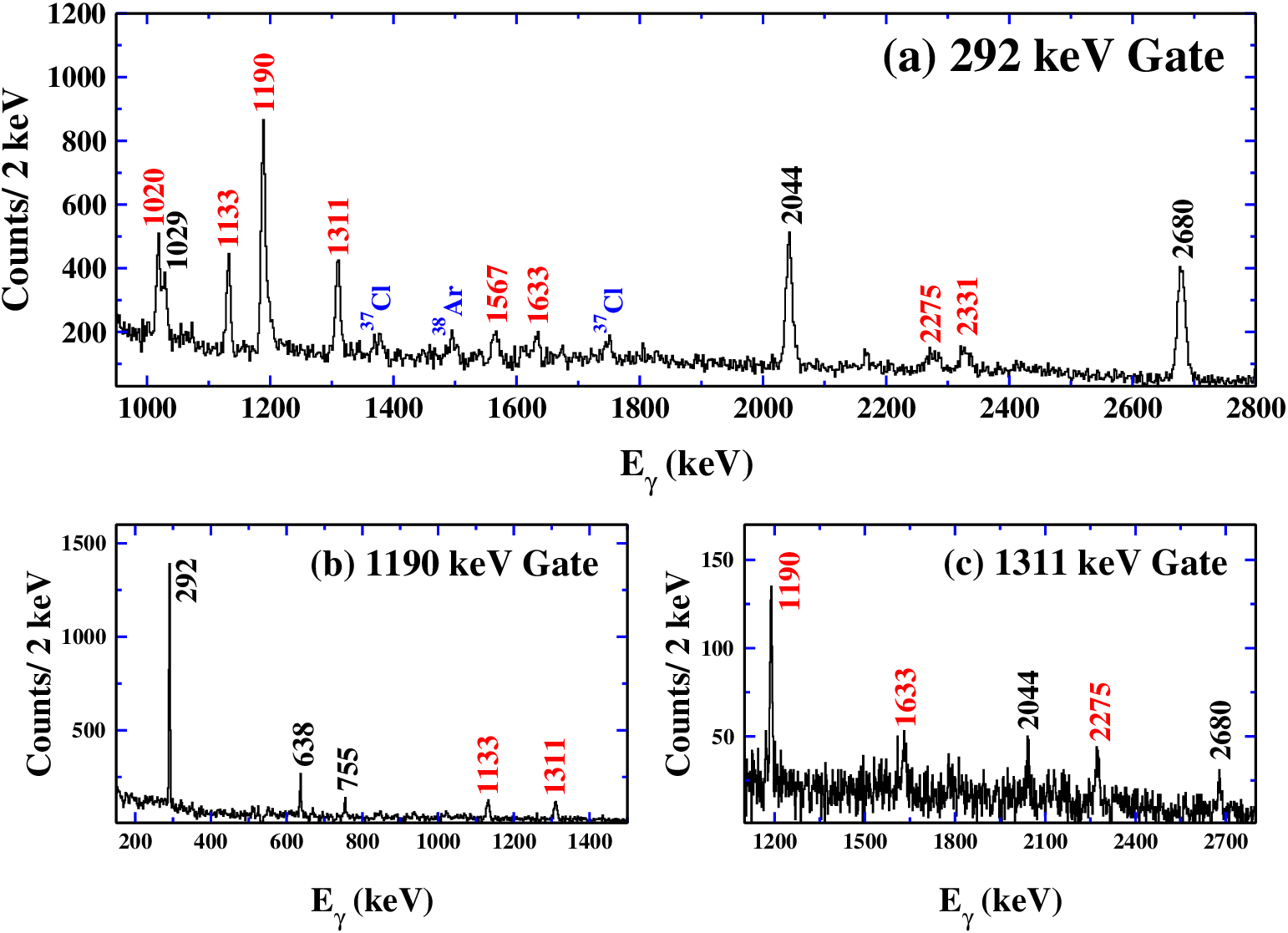}}
\caption{Doppler corrected, proton gated, $\gamma$ ray coincidence spectra. (a) Gate around the 292 keV transition shows a number of new $\gamma$ transitions. The new $\gamma$ peaks are in red (gray).  (b)(c) Reverse gated spectra confirming the new transitions. The peaks not labeled as belong to $^{38}$Cl in (a) come from random or background coincidences with strong $^{38}$Ar and $^{37}$Cl lines. }
\label{fig:292_gate}
\end{figure}

The level scheme of $^{38}$Cl as deduced from the present work is shown in Fig. \ref{fig:38_level}. Different gates  taken from the proton - $\gamma$ - $\gamma$ coincidence matrix and consideration of the observed intensity patterns helped us to add eleven new $\gamma$ transitions to the level structure, leading to six new states of $^{38}$Cl. The level scheme was extended up to 8420 keV energy, along with spin-parity assignments to a number of states. Figure \ref{fig:292_gate} shows proton gated $\gamma - \gamma$ coincidence spectra confirming a number of the new $\gamma$ transitions. The measured  $\gamma$-ray energies, multipolarities of the transitions, and suggested spins and parities are summarized in Table \ref{tab:38Cl_ex}.   \\

\begin{figure}[h!]
\centerline{\includegraphics[scale=0.7]{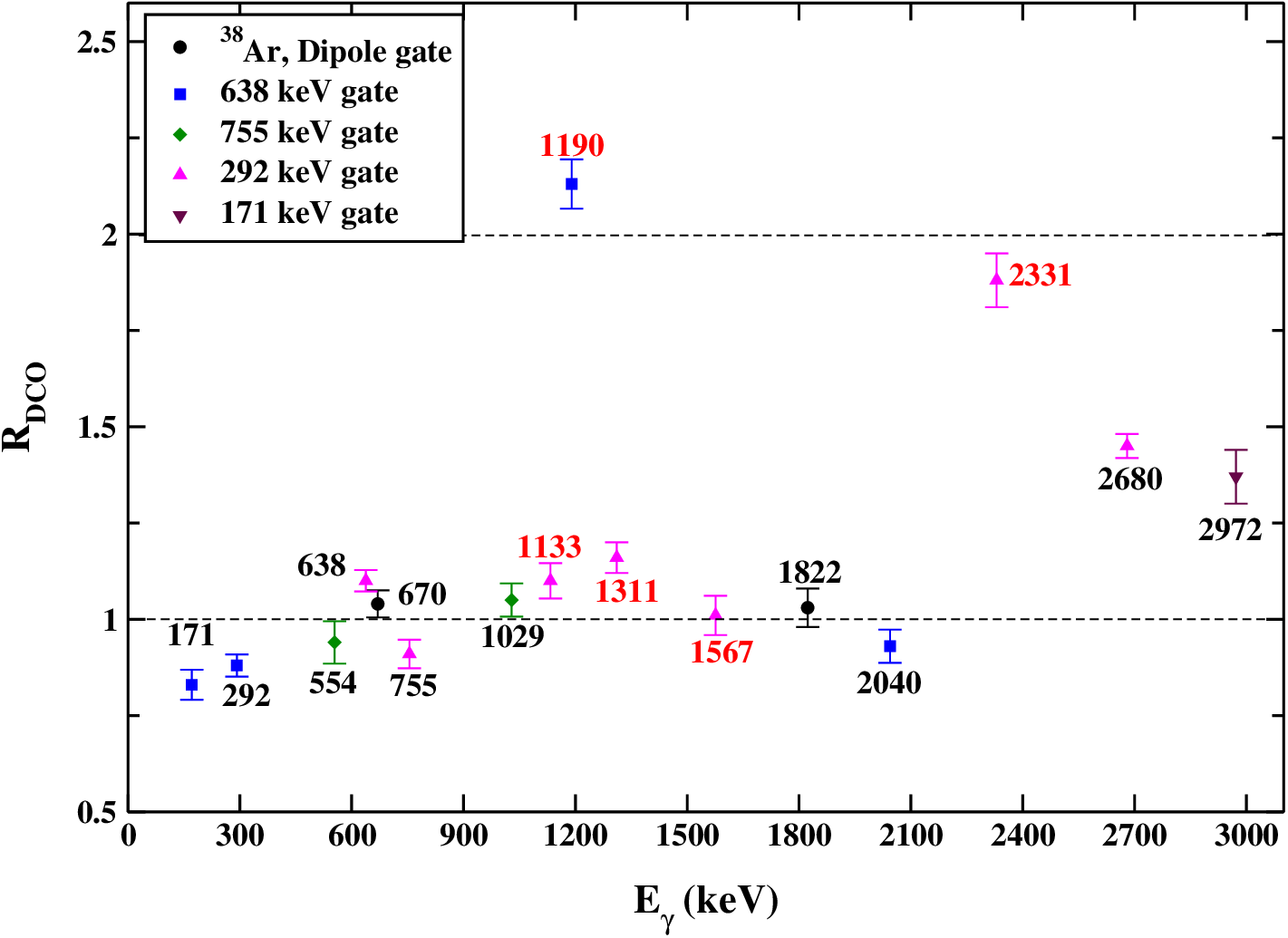}}
\caption{The experimental $R_{DCO}$ values for $\gamma$ transitions of $^{38}$Ar and $^{38}$Cl isotopes in different gates. Since the multipolarity of the transitions of 670 (M1) and 1822 (E1) keV of $^{38}$Ar are well established \cite{nndc}, we have presented the measured $R_{DCO}$ as a test case.}
\label{fig:RDCO}
\end{figure}

The lowest $2^-$, $5^-$, $3^-$, $4^-$ multiplet, resulting primarily from the $\pi (d_{3/2})^1 \otimes \nu (f_{7/2})^1$ configuration, was observed in most of the previous studies and also verified in the current work. The isomeric transition from the 671-keV, 5$^-$ state ($t_{1/2}$=715) to the ground state was not observed in the current analysis, because the time gate for the coincident events in the current experiment of 100 ns.  However, we observed a number of transitions decaying to this state because it is the first excited state and has the highest spin below 3300 keV. Two more previously known non-yrast states at 1617 and 1784 keV were also observed with the confirmation of three $\gamma$ transitions from them. No spin assignment had been made to the 1784 keV state earlier. The current data allowed us to measure the DCO ratio of the 1029 keV transition decaying from this state. The $R_{DCO}$ value measured in the dipole 755-keV gate is 1.05 (4) (Fig. \ref{fig:RDCO}), which indicates the 1029-keV transition has multipolarity 1. Since the state decays to a 3$^-$ level, we suggest spin 4 following the nature of fusion-evaporation reactions of populating higher spin states.\\

\begin{figure}[h!]
\centerline{\includegraphics[scale=0.7]{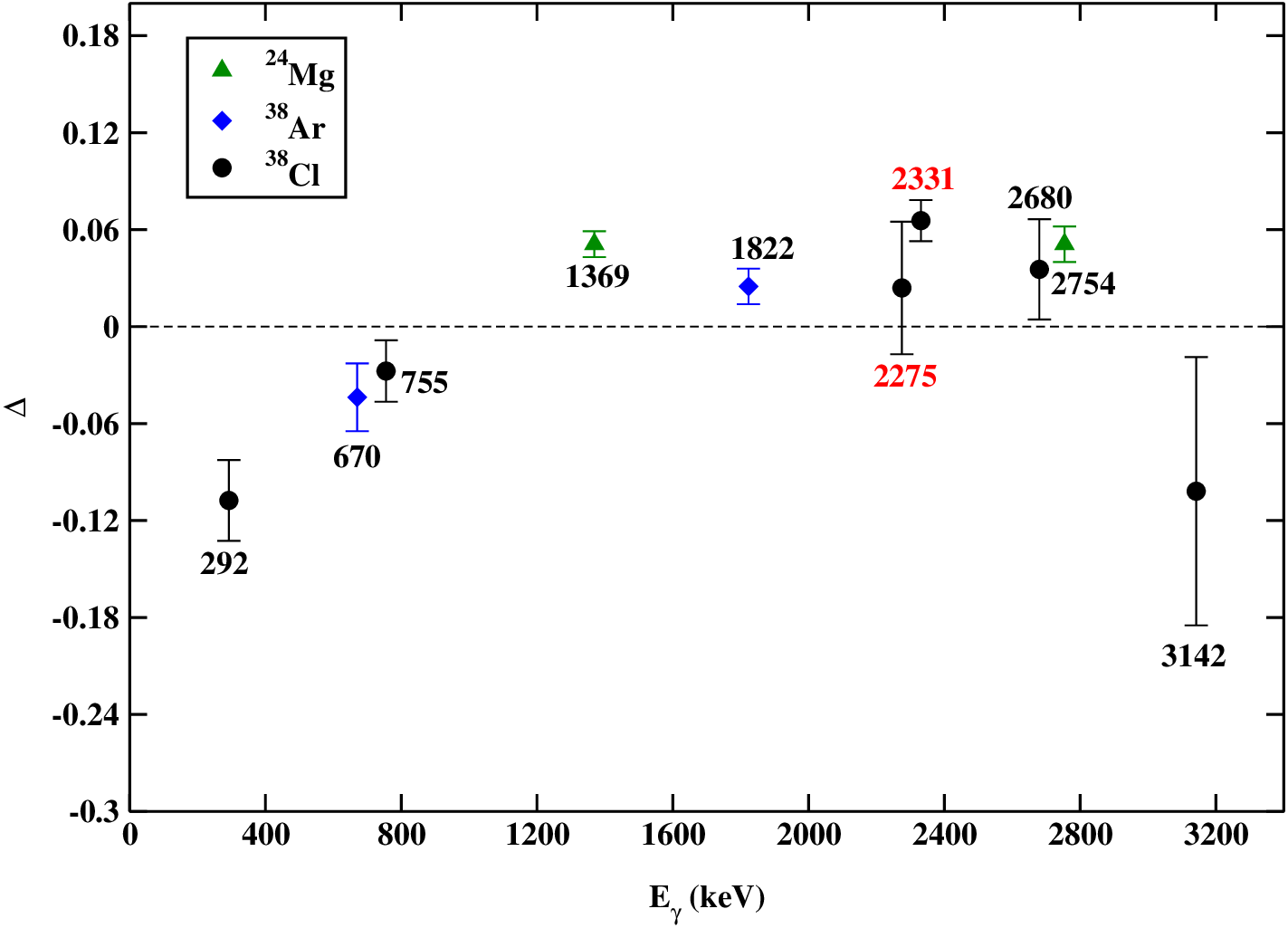}}
\caption{Polarization asymmetry of $\gamma$ transitions in $^{38}$Cl, $^{24}$Mg, and $^{38}$Ar. $^{24}$Mg points are included in the figure to show the polarization sensitivity of the clover detectors.  The two transitions in $^{38}$Ar at energies of 670 and 1822 keV are magnetic and electric in nature, respectively, according to the literature \cite{nndc} verifying the current analysis. }
\label{fig:polarization}
\end{figure}

In Ref. \cite{heavyIon1}, O' Donnell $et \,al.$ discussed possible assignments of 7$^+$, 6$^+$, and 5$^+$ to the states at energies of 3349, 3639 and 3809 keV, generated from the $(f_{7/2})^2$ configurations, though no experimental analysis was presented to assign spins or parities. We have observed these 3 states at slightly different energies of 3352, 3643 and 3814 keV. The $R_{DCO}$ measurement gated from the well-established dipole 638-keV transition implies that the transition at 2044 keV is of stretched dipole type, as shown in Fig. \ref{fig:RDCO}. The transition at 2680 keV was not observed in any $\gamma$-ray gate of previously known multipolarity in the current data. Therefore, we measured $R_{DCO}$ of the 292 keV $\gamma$ from the 638-keV gate, and found that the 292-keV line is a dipole type transition. Then we used the 292 keV transition as a gate for the 2680 keV $\gamma$ to measure its $R_{DCO}$. The same procedure was followed for the transitions which are not in coincidence with any $\gamma$ of known multipolarity. However, the DCO ratio of the strong transition at 2680 keV looks indecisive while gated from the 292-keV transition. The electric character of the 2680-keV transition was confirmed by measuring its polarization asymmetry, as displayed in Fig. \ref{fig:polarization}. \\

Combining the $R_{DCO}$ of the 2044-and 2680-keV transitions along with the polarization asymmetry measurement of the 2680 keV line, we propose spin-parity of $5^+$ to the 3352 keV level. According to this $J^\pi$ assignment the $\gamma$ at 2680 keV is a non-stretched transition and that might be the reason for its anomalous $R_{DCO}$ value. Also, a new transition at 1567 keV was observed from the 3352-keV level decaying to the 1784-keV state. The measured $R_{DCO}$ value of the 1567-kev $\gamma$-ray decaying from the 3352-keV level suggests that the transition is of dipole type (Fig. \ref{fig:RDCO}). Hence, it is clear that the spin of the previously observed 1784-keV level is 4 $\hbar$ as proposed before. The $R_{DCO}$ values for the 292- and 171-keV transitions are 0.88(3) and 0.83(4), respectively, while measured  in the dipole gate at 638 keV (Fig. \ref{fig:RDCO}), indicating dipole type. Also, both transitions are magnetic in character according to the measured  polarization asymmetry (Fig. \ref{fig:polarization}), which means that the 292-and 171-keV decays are of M1 type. Following the tendency of fusion-evaporation reactions to favor higher spin states, there are strong indications that the 3643- and 3814-keV levels are the yrast $6^+$ and $7^+$ states, respectively. These assignments agree with those in Ref. \cite{heavyIon1} that all three are positive parity and higher spin states. However, we suggest a change of spin sequence in concordance with the DCO ratios and polarization asymmetry measurements. Confirmation of the previous suggestion of positive parity to these 3 states is significant because it implies that they are intruder states with one additional particle promoted across the $N = 20$ shell gap having parity opposite to the ground state, just the type of states we want to study further. \\

This rich data set and the tendency of heavy-ion fusion-evaporation reactions to favor higher spin states allowed considerable further exploration of the structure of $^{38}$Cl.  A new excited state at 4587 keV was built with the observation of the 733-keV transition which decays to the 3814-keV $7^+$ level. No $R_{DCO}$ or polarization asymmetry measurement was possible for this weak transition. Two newly observed $\gamma$ transitions at 1020 and 1190 keV decaying to the 3814- and 3643-keV states, respectively, imply a new state at 4833 keV. The DCO ratio analysis of the 1190-keV transition suggests it is of quadrupole type. No $R_{DCO}$ measurement was possible for the 1020-keV $\gamma$-ray since this is a weak transition and not adequately resolved from the 1029-keV line. Since both transitions from this level decay to the highest spin states previously known, 7$^+$ and 6$^+$, and transitions are seen from even higher in the level scheme, the 4833-keV state must have higher, rather than lower spin, leading to an assignment of $8\,\hbar$.  No polarization measurement was possible for any of these transitions. However, since the level decays to the positive parity and higher spin states, it is very likely to be the yrast (lowest state of a given spin) $8^+$ state. A new level at 5966 keV was built with the observation of a $\gamma$ transition at 1133 keV. Judicious gating helped to place this line above the 4833-keV state. The transition is of dipole type according to the $R_{DCO}$ value (Fig. \ref{fig:RDCO}). Because no other transition was observed from this state and the polarization of the emitted $\gamma$-ray could not be measured, we propose any of the spins $7,\, 8,\, \text{or}\, 9\,\hbar$ for the state.  Another new level at 6145 keV was observed to emit two $\gamma$ rays of energies 1311 and 2331 keV. The decay patterns and the measured multipolarities of the associated $\gamma$ transitions suggest spin  $9\,\hbar$ for this level. Positive parity was confirmed by observing the electric type character of the 2331-keV transition, Fig. \ref{fig:polarization}.  Two new $\gamma$ transitions at 1633 and 1816 keV are found to decay from the 7779-keV level. No DCO ratio or polarization measurement was possible for either of these transitions because of insufficient statistics, thus leaving the state with unassigned spin and parity. The highest energy state observed in the current analysis is at 8420 keV and decays to the 6145-keV level via a 2275-keV $\gamma$ transition. The polarization asymmetry shows that $\Delta$ for the transition is very close to zero and it can be either electric or magnetic type within the error bar, (see Fig. \ref{fig:polarization}). However, the intensity of the transition at 2275 keV is higher than that decay from the 7779-keV state. Hence, it is more probable that the 8420-keV state is an yrast state instead of the 7779-keV state. There will be more discussions on these states in the theory section.\\

\begin{figure}[h!]
\centerline{\includegraphics[scale=0.7]{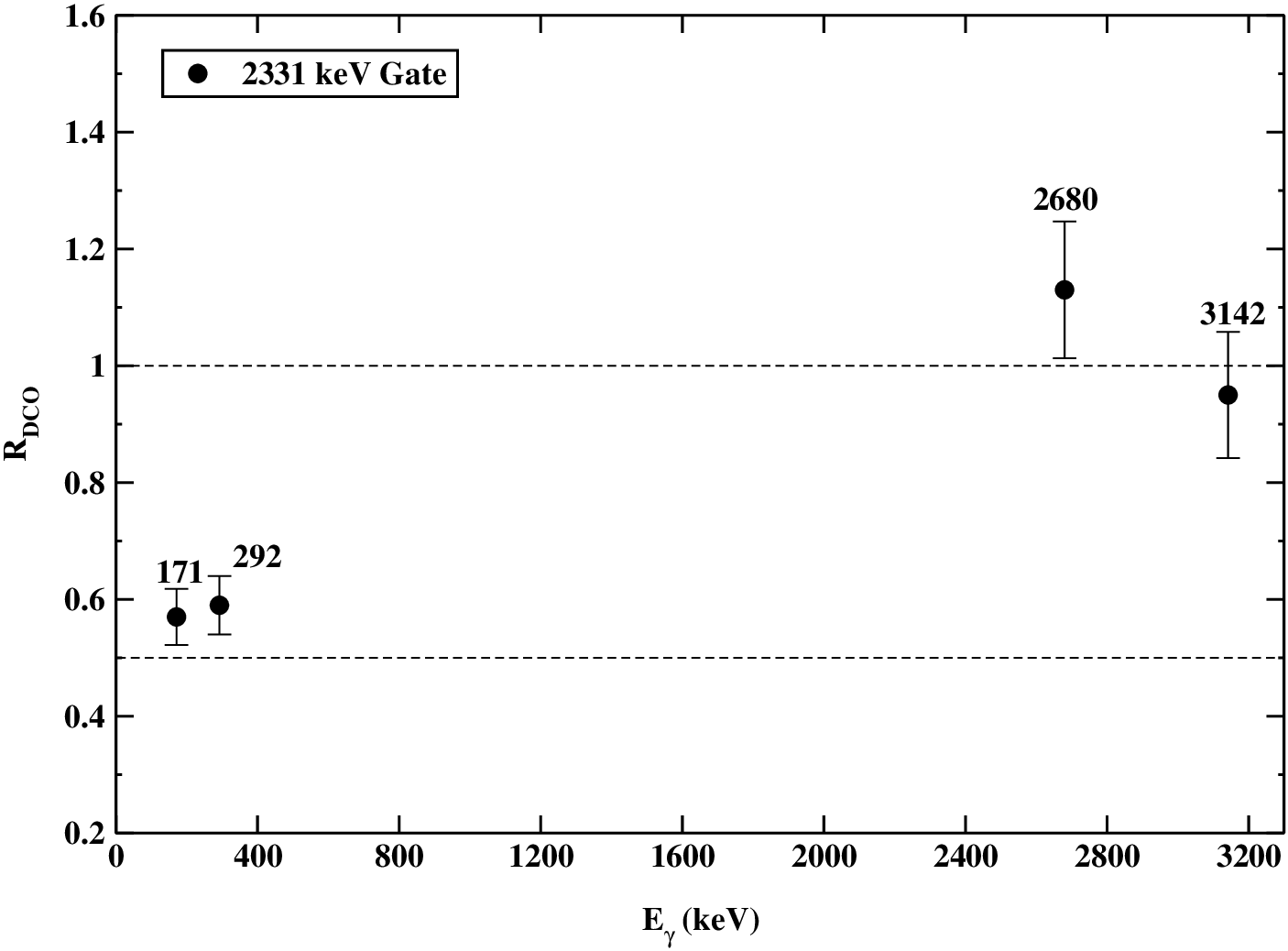}}
\caption{Experimental $R_{DCO}$ values for $\gamma$ transitions of $^{38}$Cl in the 2331 keV gate. In Fig. \ref{fig:RDCO} it is shown that 2331-keV $\gamma$ transition is of quadrupole type. In this gate, the $R_{DCO}$ values for diopole transitions are $\sim$ 0.5 and close to 1 for quadrupole transitions.}
\label{fig:RDCO2331}
\end{figure}

As seen in Fig. \ref{fig:RDCO}, the 2331-keV $\gamma$ decaying from the 6145-keV state is of quadrupole type. In the 2331-keV gate, we were able to see $\gamma$ transitions decaying from the 3643- and 3814-keV states very clearly and could measure $R_{DCO}$ of the transitions from this gate  as displayed in Fig. \ref{fig:RDCO2331}. $R_{DCO}$ values for the 292- and 171-keV $\gamma$ transitions were measured as  0.59 (5) and 0.57 (5) in the 2331 keV gate , meaning that they are of dipole type which are consistent with those measured in the 638 keV gate. A $R_{DCO}$ measurement in the 2331-keV gate was made for the previously known transition at 3142 keV, decaying from the 3814-keV level. The value is close to 1 and, because it was from a quadrupole gate, we can conclude that the transition is also of quadrupole type. The polarization asymmetry measurement as shown in Fig. \ref{fig:polarization} suggests the transition is of magnetic type. The stretched M2 multipolarity of the 3142-keV  $\gamma$ transition provides added confidence in the $J^{\pi}$ assignment of $7^+$ to the 3814 keV state.\\ 
 

\section{The FSU effective interaction for the $spsdfp$ shell model space}

The nuclear shell model studies have been successfully carried out for a long time with gradually expanded reach \cite{caurier:2005a, brown:2001}. Studies where the valence space is restricted to a single major harmonic oscillator shell have been done in the lower $p, sd, \text{and } pf$ shells, resulting in reliable, empirically established interactions \cite{CKI, usd, gxpf1a} with a relatively small number of parameters and achieving a root-mean-squared (rms) deviation from experimental data of just a couple hundred keV.  \\

The universal $sd$-shell interaction (USD) \cite{usd} has been very successful in describing the natural-parity states with $0 \hbar \omega$ configurations of the $sd$-shell nuclei.  With the availability of more experimental data, the interaction was refitted to its more recent versions, called USDA and USDB \cite{usd_ab}. The unnatural parity states, originating from cross-shell excitations of the nucleons are important in order to understand the shell gap evolution within the $sd$-shell nuclei. The shell model description for the cross-shell excitations, giving rise to the intruder states, requires the inclusion of the $0p$ and $0f$ - $1p$ shells, which is beyond the model space for which the USD family of interactions apply.  As the $sd$ shell is surrounded by two major shells, $0p$ and $0f$ - $1p$ from the lower and upper ends respectively, the intruder states of the $sd$ isotopes can involve two types of configurations, $0p^{-1}(1s0d)^{1}$ more prevalent lower in the $sd$ shell and $(1s0d)^{-1}(0f1p)^{1}$ (relative to the g.s. configuration) more prevalent higher in the shell and, of course, combinations of both structures in the same states. Hence, theoretical predictions of the intruder states of $sd$-shell nuclei require the $psdfp$ model space or, at least, $sdfp$ when the focus is more on higher mass isotopes. A number of theoretical efforts have been made recently to interpret the intruder states and to understand the $sd$-$fp$ shell gap evolution with increasing neutron number. As an example, the SDPF-M interaction \cite{sdpf-m} has been successful in reproducing the excited states of lower $Z$ neutron-rich even-$Z$-even-$N$ $sd$ shell nuclei, but was unable to reproduce the ordering and the energies of the isotopes with odd $N$, as mentioned in Ref. \cite{wbp-m}. The WBP interaction \cite{wbp} was developed earlier to address cross-shell structure around $A\sim 20$. Some attempts have been taken to modify the WBP interaction \cite{wbp} by reducing the single particle energies (SPEs) of the $fp$ shell orbitals to reproduce the states of some specific isotopes of interest. For example, the SPEs of the $0f_{7/2}$ and $1p_{3/2}$ orbitals were reduced in the WBP-A \cite{wbp-a} interaction to explain the intruder states of $^{34}$P \cite{wbp-a}. The SPEs for $0f_{7/2}$, $1p_{3/2}$ and $1p_{1/2}$ were further adjusted from WBP-A to WBP-B to predict the intruder states of $^{31}$Si \cite{wbp-b}. In another version, called WBP-M \cite{wbp-m}, all the SPEs of $fp$ shell orbitals were changed in order to reproduce the energies and the ordering of the $3/2^-$ and $7/2^-$ states $^{27}$Ne which eventually fixed the ordering of the same states in $^{25}$Ne and $^{29}$Mg. In that paper the authors mentioned the need of an interaction that will reduce ``the  effective  gap  between  the  $0d_{3/2}$ orbital  and the  $0f-1p$ shell  in  a  natural  way,  without  the  need  for $ad\, hoc$ changes" . Another interaction SDPF-U \cite{sdpf-u} was developed to calculate the $0\hbar\omega$ states within the $sd-fp$ valence space. Therefore, while the interaction was quite successful in calculating the $0p0h$ states of $^{38}$Cl, it was unable to predict the intruder states as discussed by O' Donnell $et \, al,$ Ref. \cite{heavyIon1}. The PSDPF interaction spanning the $psdfp$ model space is another recent effort \cite{psdpf} to predict the intruder states of most of the $sd$-shell isotopes quite well except for those in the middle of the shell, since their intruder states were not fitted as mentioned by the authors. Therefore, there remains a need for an interaction which would successfully predict the intruder states of the $sd$-shell nuclei, including the ones with neutrons in the $fp$ shell. This need was the driving force to building a more comprehensive effective shell model interaction at FSU essential to explain our current data on $^{38}$Cl and future works.\\

Our goal was to build an interaction for a (0+1)$\hbar\omega$ space including one-particle one-hole ($1p1h$) excitations across the oscillator shells. This choice allows for a systematic description of both natural and unnatural parity states (parity opposite to that of the ground state) while ensuring the removal of spurious center-of-mass states. Furthermore, one should be careful in selecting experimental data to fit, in order to avoid excited states that cannot be properly described within the model space. The inclusion of $1p1h$ states in particular makes the fit sensitive to the cross-shell matrix elements, a previously poorly constrained component of the interaction, and the main focus of this fit. \\

As a starting point for the fitting procedure we used a modified version of the WBP interaction \cite{wbp} which has also been applied in the past with relative success in this mass region for lower $A$. The model space for WBP interaction consists of four major oscillator shells, $0s, \ 0p, \, 1s0d, \, 0f1p$, though no modification was made related to $0s$ shell in this work. The initial modifications consisted of replacing the $sd$-shell single particle energies (SPEs) and two body matrix elements (TBMEs) with those of the USDB interaction \cite{usd_ab}, and the $fp$ shell  ones with those of the GXPF1A interaction \cite{gxpf1a}. This further reduced the deviation from experiment for all $sd$-shell natural parity states and gave a somewhat more realistic description of the $fp$ shell interaction. In previous publications, we found that corrections to the single particle energies where necessary in order to correctly describe spectra~\cite{wbp-a, wbp-m, wbp-b}. Hence, we focused on the SPEs, as well as the TBMEs affecting the monopole contributions to the energy.  For the rest of the discussion, the fitted values correspond to shifts from the original SPEs and TBMEs. \\

Given the extensiveness of the model space there was no hope of fitting all SPEs and TBMEs separately, and thus we had to predetermine the important contributions to the spectra on physical grounds. For the $0p$ - $sd$ cross-shell matrix elements we varied only the monopole terms, namely those terms in the Hamiltonian that are given by the product of two number operators for the single-particle orbits involved.  With 2 $0p$ orbits and 3 $sd$ orbits there are total of 6 such parameters. No SPE was changed within the $0p$ shell. To ensure that inclusion of the lower $p$-shell does not modify the binding energies and spectra of the 0$\hbar \omega$ states in $sd$ nuclei from those of USDB,  we modified the $sd$-shell single particle energies simultaneously with the monopole terms. For the $sd$ - $fp$ cross-shell interaction, since the 1$p_{1/2}$ orbital is relatively high in energy, and thus not as sensitive to our dataset, we considered only a combined monopole correction with all the $sd$-shell orbits, namely we assumed that the additional monopole interaction introduced in the Hamiltonian has the same strength between the 1$p_{1/2}$ orbital and each of the  1$s_{1/2}$, 0$d_{3/2}$ and 0$d_{5/2}$ orbits. The two different isospin channels ($T=0,1$) of this monopole were fit as separate values.  On the other hand, the 0$d_{5/2}$ orbital is deeply bound for most nuclei considered here, and, hence we considered only the monopole terms with the $1p_{3/2}$  and $0f_{7/2}$ $pf$-shell orbitals, again with different isospin channels which amounts to four additional parameters. For the rest of the $sd$ - $fp$ cross shell interaction the remaining set of matrix elements ($0f_{7/2}$ - $1s_{1/2}$, $0f_{7/2}$ - $0d_{3/2}$, $1p_{3/2}$ - $0s_{1/2}$, and $1p_{3/2}$ - $0d_{3/2}$) is more important and here we selected all multipole-multipole density terms; going beyond monopole which corresponds to scalar. Again due to the different isospin values, this amounts to 24 parameters included in the fit. 
 Finally, for the $fp$ shell we started with GXPF1A interaction \cite{gxpf1a} but since more than just monopoles were varied for the cross-shell, it was impossible to keep the results for the $fp$ shell nuclei unchanged from the original GXPF1A. To compensate for this, and given that $fp$ interactions are less established phenomenologically than those in $sd$ or $0p$ shells, we allowed the SPEs of all 4 orbitals within the $fp$ shell  and the 30 TBMEs between  $0f_{7/2}$ and $1p_{3/2}$ to vary. All the TBMEs within the $sdfp$ space are scaled with a factor of $ A^{-1/3}$.\\

\begin{figure}[h!]
\centerline{\includegraphics[scale=0.7,angle=0]{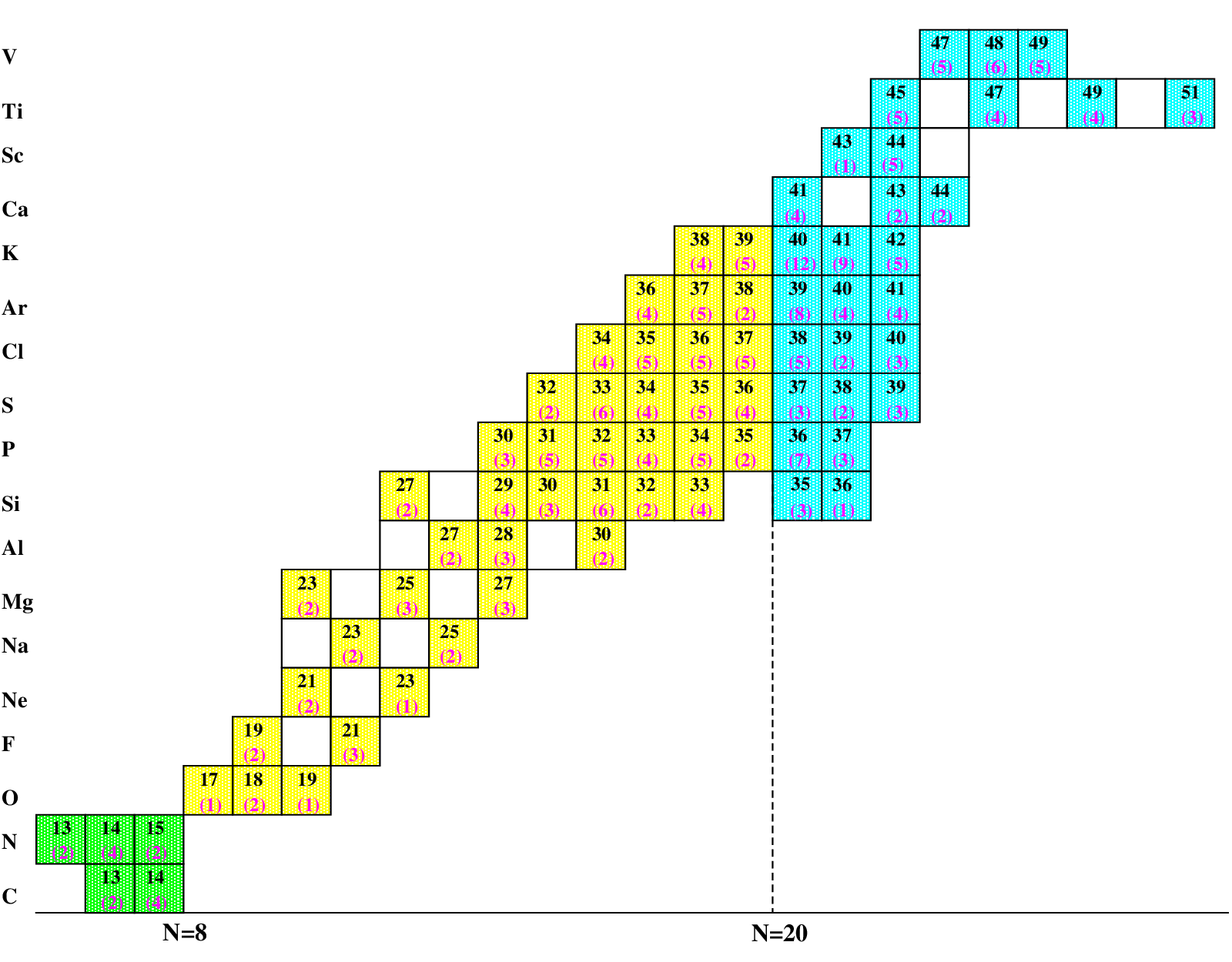}}
\caption{ Chart of the isotopes whose energy states were fitted to develop the FSU interaction. The number in each box is the mass number of an isotope and that within the parenthesis is the number of energy levels compiled for the corresponding isotope.}
\label{fig:Iso_table}
\end{figure}

The total number of varied parameters was 70 which were fitted to the 270 experimental observed states compiled from the Ref. \cite{nndc}. Out of the total 270 states, 224 are sensitive to the $N = 8$ and $N = 20$ shell gaps. These 224 states consisted of the intruder states of $0p$ and $sd$ shell isotopes along with the normal and intruder states of $sd$ shell isotopes with $N > 20$. Among these 224 energy states, the ones arise from at least two nucleons within the $fp$ shell are also sensitive to the $fp$ parameters varied. The rest 46 states are $0p0h$ states for pure $fp$ shell isotopes which were included in the fit in order to better constrain the $fp$ shell TBMEs between the $0f_{7/2}$ and $1p_{3/2}$ orbitals. The list of isotopes included in the fit is displayed in Fig. \ref{fig:Iso_table}. The minimization procedure followed the method described in \cite{usd_ab}. Given the starting Hamiltonian, we iteratively solved the eigenvalue problem for all nuclei of interest and using the resulting wave functions we linearized the energies in terms of the fit parameters and minimized the $\chi^2$ deviation from experiment resulting in new values for the fit. Not all linear combinations of the parameters were sensitive to the data chosen and some appeared to be strongly correlated. For this reason, the inverse of the error matrix defining the curvatures  in the multidimensional space near the $\chi ^2$ minimum was generally ill conditioned. We followed the standard procedure and by diagonalization of the error matrix we selected 40 linear combinations of parameters that correspond to directions in the parameter space with largest curvature near the minimum. Once the best fit parameters were determined the whole procedure was iterated starting from a new Hamiltonian. We reached the convergence after 6 iterations with an overall rms deviation from experiment of 190 keV.   \\


\section{Shell-Model Calculations and Discussions on $^{38}$Cl}

The lowest four states in $^{38}$Cl ($2^-$, $5^-$, $3^-$, and $4^-$) have long been considered to arise from the $\pi 0d_{3/2} \otimes \nu 0f_{7/2}$ configuration, although it was also realized that $\pi 0d_{3/2} \otimes \nu 1p_{3/2}$ could contribute to the $2^-$ and $3^-$ states.  The comparison of measured \cite{transfeReac2} and calculated properties of several of the lower states are shown in Table \ref{tab:negParity38Cl}.  The agreement in energy is very good, but it should be noted that the energies of the first 4 excited states were among the 270 ones used in the fit.  The calculated occupancies of the orbitals which appear to play a role in the structures are shown.  They show that the lowest $2^-$, $5^-$, $3^-$, and $4^-$ states do involve primarily the $\pi 0d_{3/2}$ and $\nu 0f_{7/2}$ orbitals, and, as suspected, there is some occupancy of $\nu 1p_{3/2}$ for the lower spin states.  Interestingly, the second $3^-$ state involves somewhat more $\nu 1p_{3/2}$ than $\nu 0f_{7/2}$.  The $\nu f_{7/2}$ and $\nu 1p_{3/2}$ occupation numbers of the second $4^-$ state are very similar to the first $4^-$ with a dominant proton excitation from the $\pi 1s_{1/2}$ to $\pi d_{3/2}$ orbital as shown in Table \ref{tab:negParity38Cl}. Fortunately, there is an excellent older measurement of $^{37}$Cl$(d,p)$ spectroscopic factors (s.f.) which tests these configurations \cite{transfeReac2}, as shown in Table \ref{tab:negParity38Cl}.  In fact, direct observation of the angular distributions reveals an $\ell = 1$ component for the $2^-$ and $3^-$ states.  For numerical comparison, the calculated s.f. using the FSU interaction are also presented. It can be seen that the predicted s.f. are consistent to those measured before \cite{transfeReac2}.  The measured and calculated s.f. for the second $4^-$ state are consistent with the $(\nu f_{7/2})^1 \otimes (\pi 1s_{1/2})^{-1}$ hole configuration which would be only weakly present in the ground state configuration of $^{37}$Cl.  A comparison of the measured lifetimes \cite{nndc} on many of these states has been presented in Table \ref{tab:38Cl_life}. The effective charges of 0.45 $e$ and 1.36 $e$ for the neutron and proton, respectively, were used as discussed in Ref \cite{usd-ab2}.  The calculations are satisfactory given the experimental uncertainties and the shell model sensitivity to the harmonic oscillator length.\\

\begin{figure}[h!]
\centerline{\includegraphics[scale=0.7,angle=0]{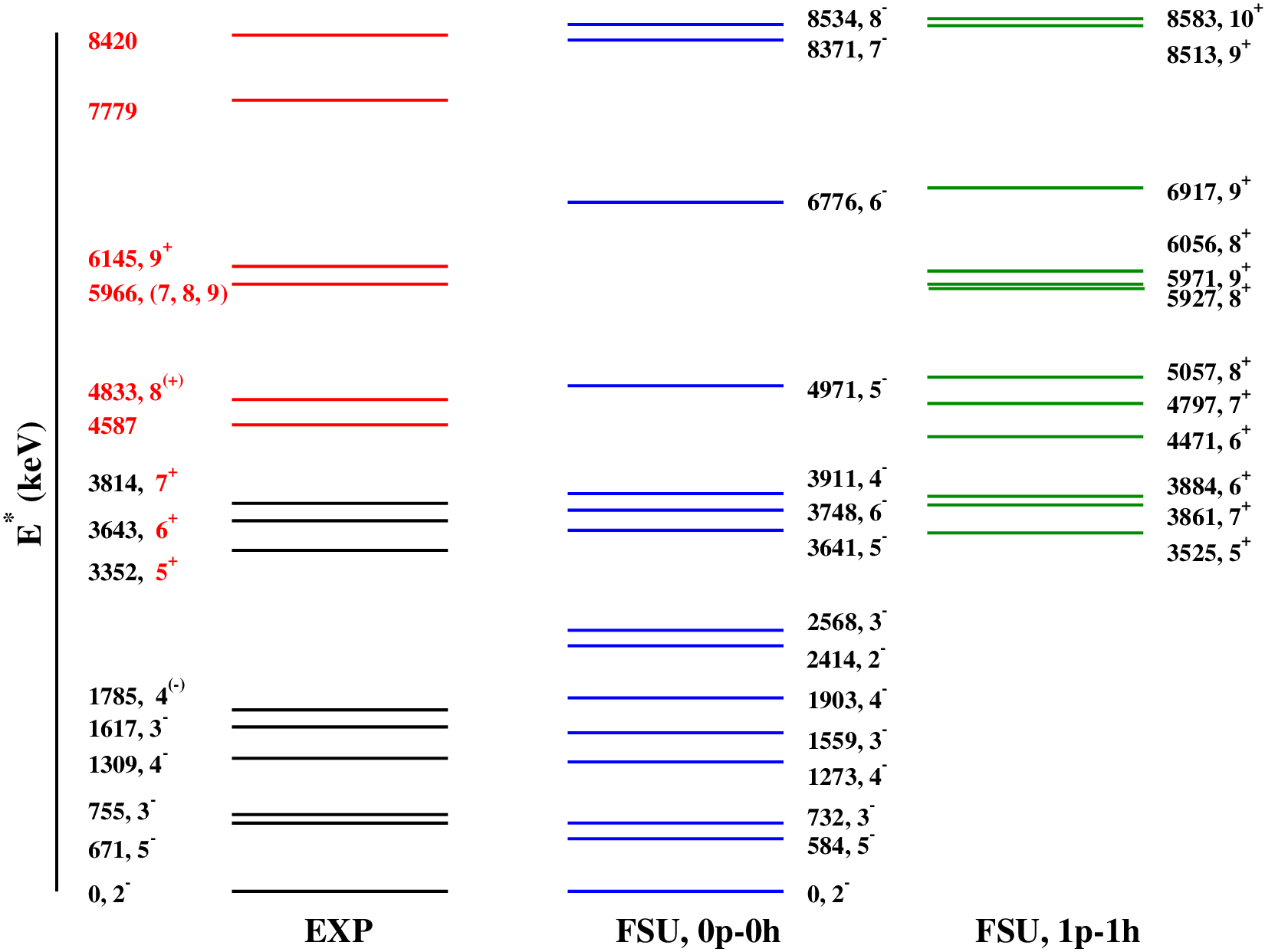}}
\caption{ Comparison of the experimental observations for $^{38}$Cl to the shell-model calculations using FSU interaction. The experimental levels in red (gray) are newly observed in the current analysis. }
\label{fig:38Cl_th}
\end{figure}

The energies of the states observed in the present experiment are compared in Fig. \ref{fig:38Cl_th} with shell model calculations using the newly developed FSU interaction for both $0p0h$ and $1p1h$ (relative to that of the g.s.) configurations.  Table \ref{tab:posParity38Cl} shows the energies and occupancies for selected $1p1h$ and $2p2h$ states calculated with the FSU interaction.  It can be seen that there are good theoretical $1p1h$ candidates for the experimental $5^+$, $6^+$, and $7^+$ closely spaced triplet, although not with the same ordering.  Interestingly, there is a triplet of $0p0h$ $\pi$ = - states predicted in the same energy range (see Fig. \ref{fig:38Cl_th}), but with a $4^-$ instead of $7^+$ state, and the $6^-$ state actually lies 127 keV below the $6^+$ one.  However, the predicted $0p0h$ $7^-$ state lies at 8357 keV, showing that the yrast sequence passes from $0p0h$ to $1p1h$ between 6 and 7 $\hbar$.  This big jump of 4.6 MeV from $6^-$ to $7^-$ results from the high energy cost of promoting a particle out of the deep 0$d_{5/2}$ orbital which substantially exceeds the energy ``cost'' of promoting one from 0$d_{3/2}$ to 0$f_{7/2}$.  A similar result has been seen in calculations for other nuclei in the upper $sd$ shell.  The strong tendency of heavy-ion fusion-evaporation reactions to populate yrast or near yrast states means that the newly observed, higher-spin states will not have $0p0h$.  Because we observe states fed by decay of higher-lying, higher-spin states and parity-changing electromagnetic transitions are generally weaker, this tendency favors populating even the non-yrast $5^+$ and $6^+$ state before the $\gamma$ decay strength flows into the $0p0h$ states.  Table \ref {tab:posParity38Cl} shows that all three $\pi = +$ states have significant $\pi 0f_{7/2}$ occupancy with the highest being 71$\%$ for the lowest $7^+$ state.  Promotion of a proton somewhat lower in the shell is significant because an $f_{7/2}$ proton and neutron can couple to $7 \hbar$, whereas the somewhat less energy costly $(\nu f_{7/2})^2$ configuration can only achieve spin $6 \hbar$.  Interestingly, most of the extra neutron excitation in the $5^+$ state goes to 1$p_{3/2}$ rather than 0$f_{7/2}$, implying a closeness of these two intruder orbitals.    \\

For the states observed above 4 MeV, candidates with both $1p1h$ and $2p2h$ excitations were considered in the calculations. The likely correspondences between the experimentally observed states above 4 MeV and the predicted states are shown in Table \ref{tab:posParity38Cl}. These correspondences are consistent with the experimental spins and parities, decay modes, and the systematics of fusion-evaporation reactions.  The calculated states  suggested to correspond to the experimental ones are the lowest and next to lowest in energy of each spin.  There are good candidates for the observed states at 4587, 4833, 5966 and 6145 keV. The predictions support the suggested spins and parities discussed in the analysis section. In particular, we suggest to label the states at 4833 and 6145 keV as the yrast 8$^+$ and 9$^+$ (see Table \ref{tab:posParity38Cl} and Fig. \ref{fig:38Cl_th}).  There is a good energy match between the observed 8420-keV state and the lowest calculated $10^+$ state. However, the predicted second and third $9^+$ states are below and above the measured 7779-keV state by over 700 keV.  For this reason, we also calculated possible correspondences involving $2p2h$ configurations.  There are good energy matches for the highest two experimental states with the predicted $9^-$ and $10^-$ $2p2h$ states, as shown in Table \ref{tab:posParity38Cl} and Fig. \ref{fig:38Cl_th}.  The lowest $1p1h$ $11^+$ and $2p2h$ $11^-$ states lie at 11419 and 9366 keV, respectively, showing that neither is a candidate for the experimental 8420 keV state.  It is quite possible that unseen $2p2h$ states above 9 MeV feed preferentially $9^-$ and $10^-$ states at 7779 and 8420 keV as seen for the $5^+,6^+, \text{and } 7^+$ triplet, but $1p1h$ configurations and positive parity cannot be ruled out with the current data.  The lack of a visible E2 decay from the 8420 keV state provides circumstantial support for a $10^-$ $2p2h$ assignment. \\

\section{Some properties of $^{33}$P explained by the FSU interaction}

\begin{figure}[h!]
\centerline{\includegraphics[scale=0.7,angle=0]{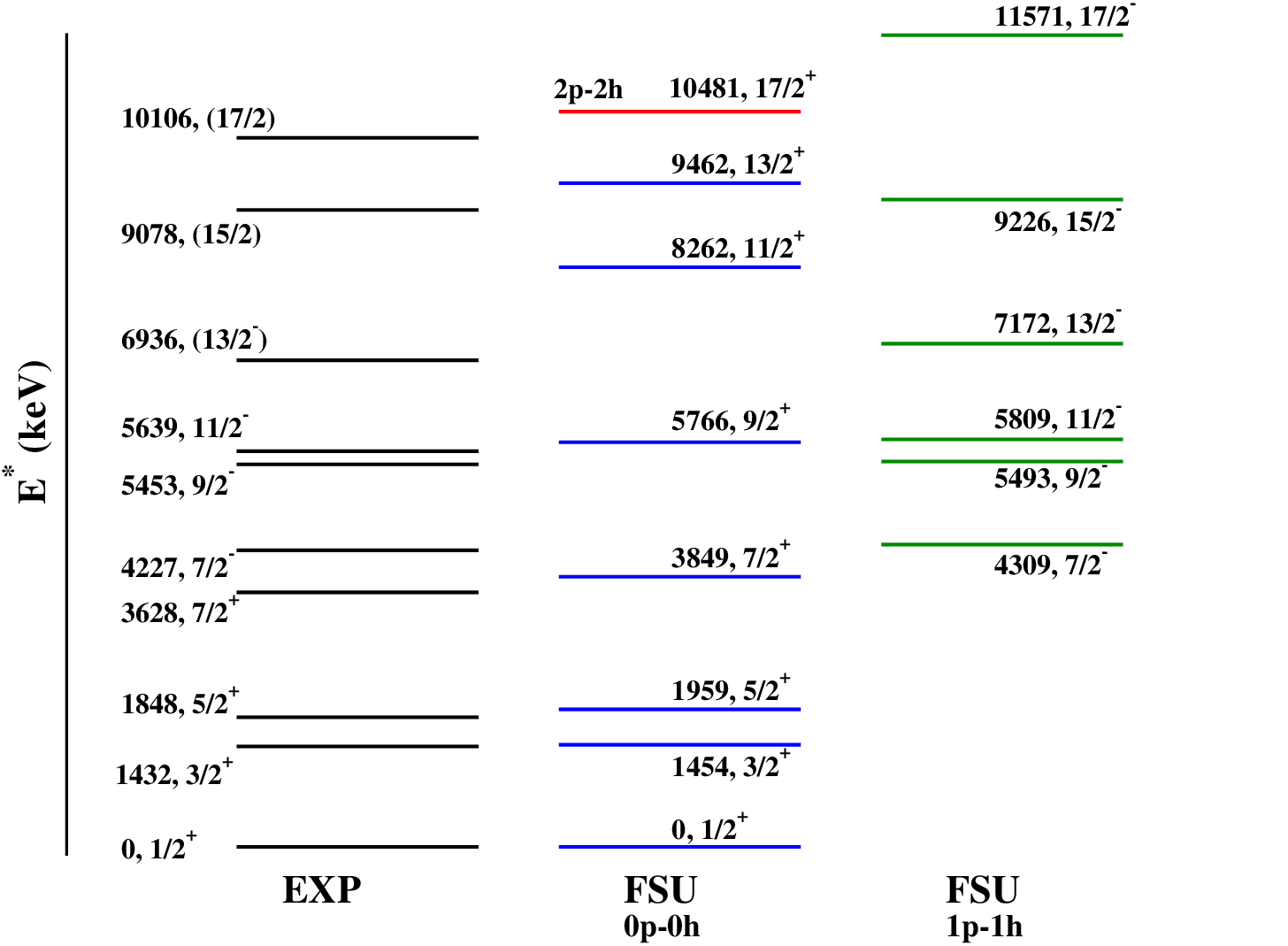}}
\caption{ Yrast states of $^{33}$P (Ref. \cite{Lubna}) are compared to the results of shell model calculations using the FSU interaction. Note that the 10481 $17/2^+$ state is calculated for a $2p2h$ configuration. The lowest $0p0h$ $17/2^+$ is calculated to lie at 15288 keV.}
\label{fig:33P_FSU}
\end{figure}

In our previous publication on $^{33}$P \cite{Lubna}, the structure was compared with shell model calculations using the PSDPF \cite{psdpf} interaction.  Although PSDPF was rather successful in reproducing the observed states, it did not predict a suitable candidate for the suggested spin $17/2$ level at 10,106 keV. We have now calculated excited states of $^{33}$P using the FSU interaction which are shown in Fig. \ref{fig:33P_FSU}. The low-energy positive-parity states calculated by the PSDPF interaction have almost the same energies as those calculated using the FSU interaction, because the $sd$ part of both are based on USDB \cite{usd_ab}. The predicted $1 \, \hbar \omega$ intruder states are in good agreement with the experimental observations, but the states from 4227 through 6936 keV were included in the fit. The positions of the $15/2^-$ and $17/2^-$ levels are true predictions.  The former lies within 150 keV of the suggested $15/2$ state, but the predicted $1p1h$ $17/2^-$ state is almost 1.5 MeV higher than the experimental 10,106 keV state.  The lowest $0p0h$ $17/2^+$ state is predicted 5 MeV higher at 15,288 keV.  A $2p2h$ configuration was suggested for this state \cite{Lubna}, but that was beyond the capability of the PSDPF interaction used in that work.  We have now calculated the lowest $17/2^+$ state using the FSU interaction which was not fitted to any $2p2h$ states.  The result is 10,481 keV, 375 keV above the experimental value.  The predominant configuration is $(\pi f_{7/2})^1 \otimes (\nu f_{7/2})^1 (\nu d_{3/2})^1$ with $f_{7/2}$ occupancies of 0.88 and 1.00 for proton and neutron, respectively.  This is essentially an aligned configuration where the $f_{7/2}$ proton and neutron are coupled to $7 \hbar$, higher than the maximum spin (6$\pi$) from two $f_{7/2}$ neutrons.  The additional $3/2 \, \hbar$ of spin comes from the remaining odd $d_{3/2}$ neutron.\\

It is worth mentioning here that all the $2p2h$ calculations performed in this work are unmixed, meaning that they are pure $2p2h$ and not mixed with the $0p0h$ configurations. The configuration mixing will have an important effect when pure $0p0h$ and $2p2h$ results lie close in energy. No significant mixing is expected for the lowest $17/2^+$ state since the 0p0p and $2p2h$ ones lie 5 MeV apart. \\

Further test of the FSU interaction were made by comparing the measured lifetimes in $^{33}$P. Two relatively long half lives of 24 ps and 9.7 ps were previously reported \cite{nndc} for the 5453-keV, $9/2^-$ and 5639-keV, $11/2^-$ states in $^{33}$P, respectively.  The lifetime of the 5453-keV, $9/2^-$ level was hard to understand because it decays by a normally fast 1226-keV M1/E2 transition.  We have calculated the lifetimes of these and other states using the FSU interaction.  The results are shown in Table \ref{tab:33P_life}.  Overall agreement is reasonably good.  The B(E2) strengths are fairly typical, ranging from 1 to 8 Weisskopf units (W.u.), but the B(M1) rates are mostly very weak.  The weak M1 strengths explain some of the longer lifetimes, especially that of the 1226-keV $9/2^- \rightarrow 7/2^-$ transition.  The abnormally (even for $^{33}$P) weak B(M1) for the 1226-keV decay may arise from the very different configurations of the two states shown in Table \ref{tab:33P_struc}.  The 5453 keV state is an almost pure proton $1p1h$ excitation, while the 4227-keV state is a mixed proton-neutron excitation.\\

\section{calculations of fully aligned, intruder states of some nearby even-$A$ Chlorine Isotopes }

\begin{figure}[h!]
\centerline{\includegraphics[scale=0.7,angle=0]{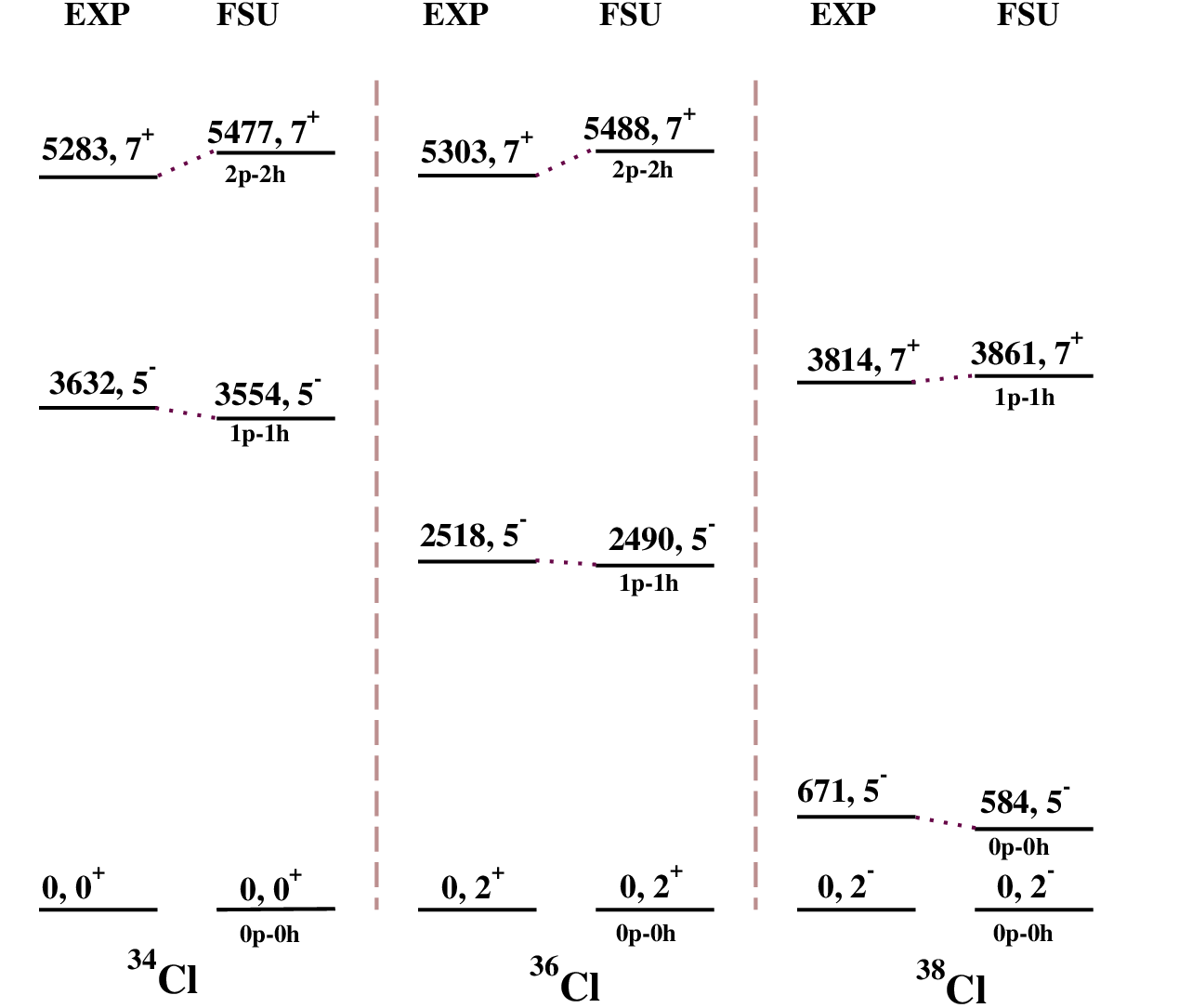}}
\caption{ Comparisons of some intruder states of even chlorine isotopes with the predicted levels employing  FSU interaction.  All of the negative parity, but none of the positive parity, excited states were included in the fit for the FSU interaction.}
\label{fig:2p-2h_Cl}
\end{figure}

It is interesting to compare the behavior of the $5^-$ and $7^+$ cross-shell states in $^{38}$Cl with the lighter, even-$A$ neighboring chlorine isotopes, as shown in  Fig. \ref{fig:2p-2h_Cl}.  The steady decrease in excitation energy of the $1p1h$ $5^-$ state with increasing $N$, even across the $N = 20$ shell gap (as it becomes a $0p0h$ excitation), is tracked rather well by shell-model calculations with the FSU interaction.  The first $7^+$ states remain fairly constant when $N < 20$ and are $2p2h$ in nature and fall significantly at $N = 21$ to become $1p1h$ in configuration.  The fully aligned $\pi f_{7/2} \otimes \nu f_{7/2}$ configuration is strongly suggested in the calculated configurations of the $7^+$ states by the nearly equal $f_{7/2}$ proton and neutron occupations shown in Table \ref{tab:even_Cl} when one realizes that most intruder states in these $N \ge Z$ states are dominated by neutron excitations. These states of $^{34, \, 36}$Cl have been selectively populated in the $^{32, \, 34}$S($\alpha, \,d$) reactions by $\ell$=6 transfer \cite{del} and, hence, confirmed to be fully aligned states.\\

\section{Summary}

Products of the reaction of a beam of $^{14}$C at 30 and 37 MeV on a  770 $\mu g/cm^2$ self-supporting $^{26}$Mg target were detected in an expanded version of the FSU $\gamma$ detector array in coincidence with an $E-\Delta E$ Si telescope at 0$^\circ$ relative to the beam. The proton-gated $\gamma$-$\gamma$ matrix allowed extending the known higher-spin level scheme of $^{38}$Cl from 3814 to 8420 keV.  Measured DCO ratios and Compton-scattering polarization asymmetries established spin-parity assignments to many of the states from $5^+$ to $9^+$.\\

The USDx family of shell-model interactions has proved valuable in understanding the structure of lower-spin states in the $sd$ shell, but not only are the maximum possible spins limited, but even the highest spins possible come at too high an energy cost compared to what is available from the $f_{7/2}$ orbital.  Also, with 21 neutrons, $^{38}$Cl lies outside the $sd$ model space.  The fact that higher-spin states in $sd$ shell nuclei necessarily involve cross-shell excitations combined with the limitations of existing multi-shell interactions, led us to undertake a new fit to the states over a wider range of nuclei from $^{13}$C to $^{49}$V and $^{51}$Ti, leading to the FSU interaction.\\

The new FSU interaction has given more confidence in our interpretation of the structure of  $^{38}$Cl.  It may not be surprising that the quadruplet of 
$\pi d_{3/2} \otimes \nu f_{7/2}$ states ($2^-$, $5^-$, $3^-$, and $4^-$) were reproduced rather well since they were included in the states fitted, but the calculated $(d,p)$ spectroscopic factors also agreed rather well with experiment.  The assignments of $5^+$, $6^+$, and $7^+$ made experimentally for the triplet of states around 3.5 MeV and the good correspondence with calculated $1p1h$ ones illustrates how the electromagnetic decay pattern can flow through a few non-yrast levels with configurations more similar to higher states before reaching the yrast ones.\\

Although the FSU interaction was fitted to $0p0h$ and mainly $1p1h$ states, we also performed unmixed $2p2h$ calculations for higher spin states. The energy of the $17/2 \, \hbar$ level in $^{33}$P did not agree well with the calculated $1p1h$ value in a previous paper \cite{Lubna}.  The energy calculated using the FSU interaction for the $1p1h$ $17/2^-$ state was also nearly 1.5 MeV too high, but that calculated for a $2p2h$ $17/2^+$ state was only 375 keV above experiment and is a more likely candidate.  Lifetime calculations with the FSU interaction for some states in $^{33}$P agree relatively well with experiment and show that an anomalously long lifetime results from a very weak B(M1) value due to configuration change in the decay.\\

The lowest $7^+$ states in a number of even $A$ Cl nuclei have been interpreted as arising from the fully aligned or stretched $\pi f_{7/2} \otimes \nu f_{7/2}$ configuration.  Although this state has a $1p1h$ configuration relative to the g.s. in $^{38}$Cl, it involves $2p2h$ excitations for $^{34, \,36}$Cl where these 7$^+$ stretched states have been observed experimentally in $(\alpha, \,d)$ reactions. The calculations with the FSU interaction reproduce the experimental energies rather well and show almost equal proton and neutron $f_{7/2}$ occupancies required by the stretched condition.\\

The FSU multi-shell interaction has worked relatively well in the cases tested here and will be used in future analyses in this mass region.\\

\begin{acknowledgements}
This work was supported by U.S. National Science Foundation under grant No. PHY-1712953 (FSU), U.S. Department of Energy, office of Science, under Award No. DE-SC-0009883 (FSU) and DE-AC05-00OR22725 (ORNL). Part of the manuscript was prepared at LLNL under Contract DE-AC52-07NA27344.
\end{acknowledgements}

%


\begin{center}
\setlength{\tabcolsep}{1em}
\begin{longtable}{l l c c c}
\caption{The excitation energies, associated $\gamma$ transitions and the measured multipolarities of $^{38}$Cl deduced in the present work. The new states and transitions are shown in boldface (normal text).  }
\label{tab:38Cl_ex}\\
\hline
$E_x$ (keV) & $E_\gamma$ (keV) & $J_i^{\pi}$   & $J_f^{\pi}$ & Mult. \\ [0.5ex] 
\hline
\endfirsthead
\multicolumn{4}{c}
{\tablename\ \thetable\ -- \textit{Continued from previous page}} \\
\hline
$E_x$ (keV) & $E_\gamma$ (keV) & $J_i^{\pi}$   & $J_f^{\pi}$ & Mult. \\ [0.5ex] 
\hline
\endhead
\hline \multicolumn{4}{r}{\textit{}} \\
\endfoot
\hline
\endlastfoot
755.1 (3)  		 & 755.1 (3)   	    				  & $3^-$		& 		$2^-$    &		M1\\  
1308.8 (5)  		 & 637.6 (1)   	    				  & $4^-$		& 		$5^-$   &		M1\\  
				 & 553.6 (2)					  & $4^-$		&		$3^-$    &		M1\\
			          & 1308.0 (1)					  & $4^-$		&		$ 2^-$   &		\\
1616.6 (8)  		 & 308.6 (8)   	    				  & $3^- $		& 		$4^-$    &		\\  
 				 & 860.7 (8)   	    			 	  & $3^-$		& 		$ 3^-$   &		\\  
1784.2 (5)  		 & 1029.1 (2)   	    				  & \boldmath$4^{(-)}$		& 		$ 3^-$    &	(E1) \\  
3351.8 (8)  		 &\textbf{1567.2 (17)}   	    			  & \boldmath$5^+$		&\boldmath$4^{(-)}$  &	\textbf{(E1)}\\  
 				 & 2043.5 (3)   	    				 & $\mathbf{5^+} $		& 		$ 4^-$    &	\textbf{E1}\\  
 				 & 2680.2 (3)   	    				  & $\mathbf{5^+} $		& 		$5^-$     &	\textbf{E1}\\  
3643.2 (10)  		 & 291.6 (3)   	    			  	  & \boldmath$6^+$		&\boldmath$5^+$   &		\textbf{M1}\\  
 				 & 2971.8 (13)   	    			 & \boldmath$6^+$		& 		$5^-$    &		\textbf{E1}\\  
3813.7 (11)      	 & 170.7 (3)   	    			 	& \boldmath$7^+$			&\boldmath$ 6^+$   &	\textbf{M1}\\  
 				 & 3142.1 (13)   	    			 & $\mathbf{7^+}$			& 		$ 5^-$   &		\textbf{M2}\\  
\textbf{4587.1 (15)}  		 &\textbf{773.4 (7)}   	    			  															&& \\    
\textbf{4833.3 (15)}  		 &\textbf{1019.5 (3)}  	    			 	& \boldmath$8^{(+)}$		&\boldmath$7^+$      &		\textbf{(M1)}\\  
 				 		 &\textbf{1190.1 (4)}  	    				 & \boldmath$8^{(+)}$		&\boldmath$ 6^+$    	 &	\textbf{(E2)}\\  
\textbf{5966.4 (17)}  		 &\textbf{1133.1 (5)}  	    			 	& \boldmath$(7, 8, 9)$		&\boldmath$ 8^{(+)}$    & \\    
\textbf{6144.7 (21)}  		 &\textbf{1311.3 (10)}   	    				& \boldmath$9^{+}$		&\boldmath$ 8^{(+)}$   &		\textbf{M1}		         \\  
 				 		 &\textbf{2331.2 (23)}   	    				& \boldmath$9^{+} $		&\boldmath$ 7^+$    	   &		\textbf{E2}	                 \\  
\textbf{7779.4 (40)}  		 &\textbf{1632.9 (34)}   	    				  														&&&  \\  
 				 		&\textbf{1816.1 (3)}   	    				  														&&&  \\
\textbf{8419.8 (3)}  		 &\textbf{2275.1 (8)}   	    			        & 		&\boldmath$ 9^{+}$    	  			   &    \\   
\end{longtable}
\end{center}

\begin{table}[]
\centering
\caption{Comparison of the experimentally observed negative parity states of $^{38}$Cl to the predicted states using FSU interaction. The measured spectroscopic factors were taken from Ref. \cite{transfeReac2}.}
\label{tab:negParity38Cl}
\begin{tabular}{|c|c|c|c|c|c|c|c|c|c|c|}
\hline
\multirow{2}{*}{J$^{\pi}$} & \multicolumn{2}{c|}{Energy (keV)} & \multicolumn{4}{c|}{Theoretical Occupancy} & \multicolumn{2}{c|}{S. F. : $\ell=1$} & \multicolumn{2}{c|}{S. F. : $\ell=3$ } \\ \cline{2-11} 
                   	 & EXP	& Th           & $\pi s_{1/2}$       & $\pi d_{3/2}$      	& $\nu f_{7/2}$       	& $\nu p_{3/2}$      & EXP \footnote{Ref \cite{transfeReac2}}  & Th  & EXP \footnotemark[1] & Th \\ \hline
2$^-$                  & 0          	& 0             & 1.89     			& 1.17			& 0.95     			& 0.04      			& 0.12      	& 0.03      & 3.60        & 4.6       \\ \hline
5$^-$                  &671       & 548        & 1.92     		 	& 1.14      			& 1.00			& 0.00 			&             	&       	 & 7.50       & 10.62       \\ \hline
3$^-$                  & 755      & 732        & 1.88     			& 1.19    			& 0.89 			& 0.11 			& 0.56      & 0.63       & 3.80       & 5.83       \\ \hline
4$^-$                 & 1309    & 1273     & 1.80     			& 1.28 			& 0.95   			& 0.04 			&      		& 		 & 5.90        & 7.54       \\ \hline
3$^-$                  & 1617    & 1559     & 1.64     			& 1.43 			& 0.40			& 0.60    			& 2.00         & 4.16      &             & 0.61       \\ \hline
4$^-$                  & 1784    & 1903     & 1.10     			& 2.01 			& 0.99   			& 0.00 			&      		& 		 & 0.94\footnote{assumed J$^\pi$ = $3^-$}       & 0.77       \\ \hline
3$^-$                  & 2748    & 2568	 & 1.27     			& 1.84			& 0.72			& 0.27    			& 1.90         & 1.86      &             & 0.004       \\ \hline
\end{tabular}
\end{table}


\begin{table}
\centering
\caption{Comparison of observed and calculated half lives of some excited states of  $^{38}$Cl. The measured half lives were taken from the  Ref. \cite{nndc}. The effective charges for proton and neutron used in the calculation are 1.36 e and 0.45 e respectively, same as discussed in Ref. \cite{usd-ab2}.}
\label{tab:38Cl_life}
 \begin{tabular}{ |c |c |c |} 
\hline
Observed $E_x$ (keV) & Measured half life (ps) \footnote{Ref. \cite{nndc}}  & Calculated half life (ps) \\ [0.5ex] 
 \hline
755  		 & 0.220 (37)   	    				  & 1.04\\  
1309  		 & 0.37 (6)   	    				  & 0.82\\  
1617  		 & 1.52 (14)   	    				  & 1.71\\  
1784  		 & 0.066 (15)  	    				  & 0.21 \\  
\hline
\end{tabular}
\end{table}

\begin{table}[]
\centering
\caption{Comparison of the experimentally observed states and the suggested intruder states calculated with the FSU interaction. The $\pi$ = + states are from $1p1h$ calculations and the $\pi$ = - are from $2p2h$ ones. Parentheses in the experimental energies indicate uncertainty in the experimental spin-parity assignments. }
\label{tab:posParity38Cl}
\begin{tabular}{|c|c|c|c|c|c|c|c|c|c|c|}
\hline
\multirow{2}{*}{$J^{\pi}$} & \multicolumn{2}{c|}{Energy (keV)} & \multicolumn{8}{c|}{Theoretical Occupancy} \\ \cline{2-11} 
 & EXP & Th & $\pi s_{1/2}$ & $\pi d_{3/2}$ & $\pi f_{7/2}$ & $\pi p_{3/2}$ & $\nu s_{1/2}$ & $\nu d_{3/2}$ & $\nu f_{7/2}$ & $\nu p_{3/2}$ \\ \hline
5$^+$ & 3352 & 3525 & 1.71 & 0.93 & 0.42 & 0.07 & 1.97 & 3.57 & 1.01 & 0.43 \\ \hline
6$^+$ & 3643 & 3884 & 1.80 & 1.00 & 0.34 & 0.00 & 1.95 & 3.48 & 1.52 & 0.05 \\ \hline
7$^+$ & 3814 & 3861 & 1.76 & 0.67 & 0.71 & 0.01 & 1.94 & 3.83 & 1.18 & 0.03 \\ \hline
6$^+$ &  & 4471 & 1.77 & 1.31 & 0.08 & 0.00 & 1.93 & 3.22 & 1.82 & 0.03 \\ \hline
7$^+$ & (4587) & 4797 & 1.64 & 1.23 & 0.26 & 0.01 & 1.78 & 3.56 & 1.58 & 0.04 \\ \hline
8$^+$ & 4833 & 5057 & 1.69 & 1.35 & 0.12 & 0.00 & 1.91 & 3.27 & 1.81 & 0.01 \\ \hline
8$^+$ & (5966) & 5927 & 1.49 & 1.29 & 0.34 & 0.00 & 1.94 & 3.48 & 1.39 & 0.17 \\ \hline
9$^+$ & 6145 & 5971 & 1.85 & 1.26 & 0.04 & 0.00 & 1.99 & 3.11 & 1.89 & 0.01 \\ \hline
8$^+$ &  & 6056 & 1.25 & 1.75 & 0.17 & 0.00 & 1.97 & 3.26 & 1.68 & 0.08 \\ \hline
9$^+$ & (7779) & 6917 & 1.06 & 1.31 & 0.78 & 0.00 & 1.99 & 3.83 & 1.17 & 0.01 \\ \hline
9$^+$ & (7779) & 8513 & 0.84 & 2.02 & 0.35 & 0.00 & 1.90 & 3.52 & 1.56 & 0.02 \\ \hline
10$^+$ & (8420) & 8583 & 1.11 & 2.09 & 0.01 & 0.00 & 1.96 & 3.11 & 1.94 & 0.01 \\ \hline
11$^+$ &  & 11413 & 1.95 & 1.46 & 0.64 & 0.00 & 2.00 & 3.68 & 1.33 & 0.00 \\ \hline
9$^-$ & (7779) & 7506 & 1.52 & 1.80 & 0.05 & 0.00 & 1.53 & 2.79 & 2.74 & 0.10 \\ \hline
10$^-$ & (8420) & 8349 & 1.30 & 1.08 & 0.91 & 0.02 & 1.88 & 3.13 & 1.86 & 0.09 \\ \hline
11$^-$ &  & 9366 & 1.43 & 0.93 & 0.89 & 0.02 & 1.92 & 3.10 & 1.88 & 0.06 \\ \hline
\end{tabular}
\end{table}


\begin{table}[]
\centering
\caption{Comparison of observed \cite{nndc} and calculated (FSU interaction) half lives of some excited states of $^{33}$P. The effective charges for proton and neutron used in the calculation are 1.36 e and 0.45 e, respectively, same as discussed in Ref. \cite{usd-ab2}.}
\begin{tabular}{|c|c|c|c|c|c|}
\hline
E$_{ex}$ (keV) & E$\gamma$ (keV) & B(M1) (mW.u.) & B(E2) (W.u.) & Overall T$_{1/2}$ (ps) & EXP T$_{1/2}$ (ps)\footnote{Ref. \cite{nndc}} \\ \hline
1432 & 1432 & 1.61 & 7.70 & 1.75 & 0.43 (7) \\ \hline
\multirow{2}{*}{1848} & 1848 & 0 & 6.11 & \multirow{2}{*}{0.81} & \multirow{2}{*}{0.77 (11)} \\ \cline{2-4}
 & 416 & 17.2 & 0.94 &  &  \\ \hline
5453 & 1226 & 0.45 & 1.52 & 15.35 & 24 (5) \\ \hline
\multirow{2}{*}{5639} & 186 & 190 & 1.78 & \multirow{2}{*}{9.94} & \multirow{2}{*}{9.7 (14)} \\ \cline{2-4}
 & 1412 & 0 & 1.20 &  &  \\ \hline
6936 & 1298 & 32.3 & 2.17 & 0.43 &  \\ \hline
\end{tabular}
\label{tab:33P_life}
\end{table}


\begin{table}
\centering
\caption{$\pi f_{7/2}$ and $\nu f_{7/2}$ occupation numbers representing the configurations of the $1 \, \hbar \omega$ $7/2^-$, $9/2^-$ and $11/2^-$ states of $^{33}$P.}
\label{tab:33P_struc}
 \begin{tabular}{| c |  c |  c |  c | } 
\hline
 EXP (keV)   & FSU (keV) & $\pi f_{7/2}$ & $\nu f_{7/2}$ \\ [0.5ex] 
 \hline
4227		& 4309		& 0.32		&0.48\\  
5453		& 5493		& 0.83		&0.09\\
5639		& 5809		& 0.90		&0.05\\
\hline
\end{tabular}
\end{table}


\begin{table}
\centering
\caption{Occupation number of $\pi f_{7/2}$ and $\nu f_{7/2}$ orbitals; representing the configurations of $7^+$ states of some even $A$ chlorine isotopes.}
\label{tab:even_Cl}
 \begin{tabular}{| c |  c |  c |  c |  c|} 
\hline
Isotopes & EXP (keV)   & FSU (keV) & $\pi f_{7/2}$ & $\nu f_{7/2}$ \\ [0.5ex] 
 \hline
$^{34}$Cl  		 &5315		& 5477		& 0.89		&0.89\\
$^{36}$Cl  		 &5313		& 5488		& 0.87		&0.98\\
$^{38}$Cl  		 &3814		& 3861		& 0.71		&1.18\\
\hline
\end{tabular}
\end{table}


\end{document}